\documentclass[12pt]{extarticle}
\usepackage{graphicx}
\usepackage{framed}
\usepackage[normalem]{ulem}
\usepackage{amsmath}
\usepackage{amsthm}
\usepackage{amssymb}
\usepackage{amsfonts}
\usepackage{enumerate}
\usepackage[utf8]{inputenc}
\usepackage[top=1 in,bottom=1in, left=1 in, right=1 in]{geometry}

\newcommand{\bb}{\begin{equation}}
\newcommand{\ee}{\end{equation}}
\newcommand{\ug}{\; = \;}

\theoremstyle{definition}

\newcommand{\vs}{\vspace*}

\title{Mathematical description of a \textit{Frozen Wave} beam after passing through a pair of convex lenses with different focal distances}
\author{Michel Zamboni-Rached$^{1*}$, Grazielle de A. Lourenço-Vittorino$^1$,\\ T. Viana de Sousa$^1$,
	Joel A. Varela Mendonça$^1$,\\ Jessyca N. Pereira$^1$ and Erasmo Recami$^{2}$\\
	\\
	$^1$ University of Campinas, Campinas, São Paulo, Brazil\\
	$^2$ On leave from
	the University of Bergamo, and INFN of Milan,\\ due to a PVE fellowship by FAPESP.\\
	$^*$Corresponding author: mzamboni@decom.fee.unicamp.br}

\date{July 30th, 2019}

\begin{document}

\maketitle

\vs{0.5 cm}

{\bf Abstract  \ --} \ In this paper, we shall provide an analytical solution describing a \textit{Frozen Wave} beam after passing through a pair of convex lenses with different focal distances.


\section{Introduction}

In this work, based on the Fresnel diffraction integral, we provide the mathematical description of a \textit{Frozen Wave} (FW) \cite{Zamboni-Rached2004} \cite{Zamboni-Rached2006} beam after passing through a pair of convex lenses with different focal distances. It is shown that while the transverse spot size of such a beam is scaled by the lens focal ratio, $f_2/f_1$, the longitudinal pattern, which can be chosen on demand, is scaled by the quadratic factor $ (f_2 /f_1)^2$. The results here presented can be important in the generation of structured light beams within very small spatial regions.



\section{Mathematical description}

\begin{figure}[!htb]
	\centering
	\includegraphics[scale=0.287]{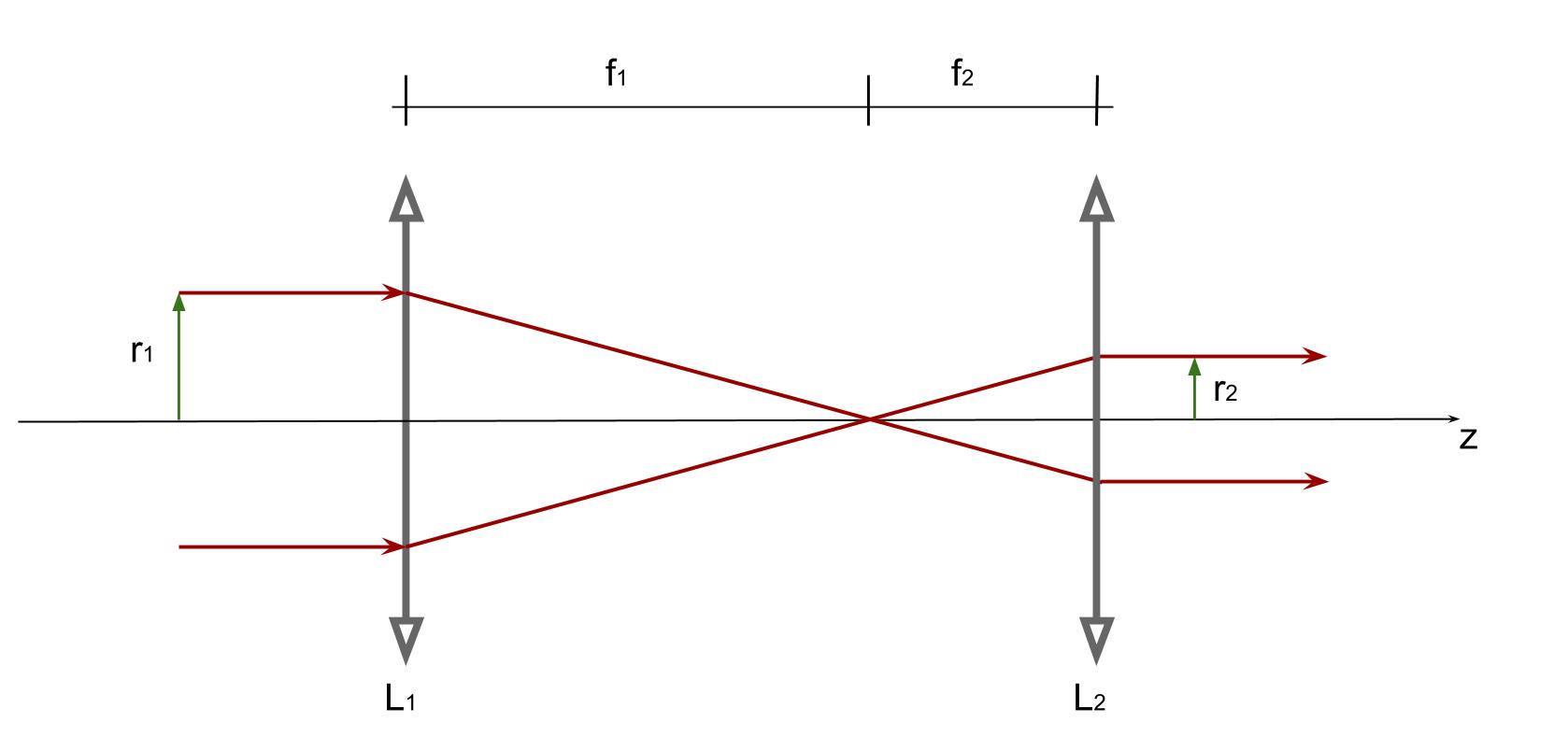}
	\caption{Sketch of a \textit{Frozen Wave} in a two-convex lens system.}
	\label{GeoOptics}
\end{figure}

From geometric optics, we know that a light beam, represented by parallel rays, with spot radius $r_1$, after passing through a system of two convex lenses, $L_1$ and $L_2$, positioned in such a way to share the same focal plane, see Fig.\ref{GeoOptics}, will acquire a spot radius

\begin{equation}
r_2 = \frac{f_2}{f_1}r_1 \,\, ,
\label{GeoOptics1}
\end{equation}
where $f_1$ and $f_2$ are the focal distances of the first and second lens, respectively.

Now, we are going to show, in a rigorous way, what happens to a FW beam when it passes through the system of two convex lenses as shown in Fig.\ref{GeoOptics}.



Let us consider a \textit{Frozen Wave}, $\Psi_{I}$, incident on the first lens, given by


\begin{equation}
\Psi_{I}(\rho,z)=\sum_{q=-N}^{N} A_q J_0(h_{Iq} \rho) e^{i\beta_{Iq}z}
\label{eq:psi1}
\end{equation}
\noindent
where $h_{Iq}$ and $\beta_{Iq}$, the transverse and longitudinal wave numbers, respectively, of the Bessel beams that constitute the FW beam, are given by

\bb h_{Iq}\ug \sqrt{2k}\sqrt{(k-Q)-\frac{2\pi}{L}q} \label{hi} \ee
and (in the paraxial approximation)

\bb \beta_{Iq} \ug k - \frac{h_{Iq}^2}{2k} \ug Q +\frac{2\pi}{L}q \,\, , \label{betai}\ee
with the coefficients $A_q$ given by

\begin{equation}
     A_q=\frac{1}{L}\int_{0}^{L}F(z)e^{-i\frac{2\pi q}{L}z}dz \,\, .
\label{eq:am}
 \end{equation}

In the above equations, $Q$ and $L$ are constants and $F(z)$ is a function chosen at will.

A beam given by Eqs.(\ref{eq:psi1},\ref{hi},\ref{betai},\ref{eq:am}) is a structured non-diffracting beam whose longitudinal intensity pattern within $0 \leq z \leq L$ given by $|F(z)|^2$, can be chosen on demand, while the spot radius, $r_1$, is approximately given by

 \begin{equation}
r_1=\frac{2.4}{h_{I0}} \,\, .
\label{eq:r1}
\end{equation}

For the interested reader, the FW theory is developed with details in \cite{Zamboni-Rached2004, Zamboni-Rached2006, Zamboni-Rached2005}.



Now, let us calculate the field after the first lens ($L_1$), but before the second one, by using the Fresnel Diffraction Integral, 

\begin{equation}
   \Psi(\rho,z)=\frac{-ik}{z}e^{ikz}e^{\frac{ik\rho^2}{2z}}\int_{0}^{\infty}\Psi(\rho',0)e^{\frac{ik\rho'^2}{2z}}J_0\left(\frac{k\rho\rho'}{z}\right)\rho'd\rho'
   \label{eq:fresnel}
\end{equation}

\noindent where, for this case, $\Psi(\rho',0)$ is the product of two terms: the incident \textit{Frozen Wave} (from Equation $(\ref{eq:psi1})$) and the lens transfer function for $L_1$ (the exponential term):

\begin{equation}
   \Psi(\rho',0)=\Psi_I(\rho',0)e^{\frac{-ik\rho'^2}{2f_1}}
   \label{eq:aux_psil1}
\end{equation}

 By using Eq. \eqref{eq:aux_psil1} in \eqref{eq:fresnel}, the field that emanates from the first lens is given by:

\begin{eqnarray}
  \Psi(\rho,z)_{l_1}=e^{ikz}\frac{-ik}{z}e^{\frac{ik\rho^2}{2z}}\sum_{q=-N}^{N}A_q\int_0^{\infty}d\rho'\rho'J_0(h_{Iq}\rho')J_0\left(\frac{k\rho\rho'}{z}\right)e^{\frac{-ik\rho'^2}{2f_1}}e^{\frac{ik\rho'^2}{2z}}
  \label{fieldl1}
\end{eqnarray}

In order to solve the integral from Equation \eqref{fieldl1}, we use the following result from \cite{Gradshteyn},

\begin{equation}
    \int_0^{\infty} x e^{-a^2x^2} J_p(\alpha x) J_p(\beta x) dx = \frac{1}{2a^2} e^{\left[-\frac{\alpha^2+\beta^2}{4a^2}\right]} I_p\left(\frac{\alpha \beta}{2a^2}\right) \,\, ,
    \label{eq:sol_int}
\end{equation}
and get


\begin{equation}
    \Psi(\rho,z)_{l_1}=-e^{ikz}\frac{1}{(z-f_1)}\sum_{q=-N}^{N} A_q f_1e^{\frac{ik\rho^2}{2(z-f_1)}}e^{i\frac{h_{Iq}^2f_1}{2k(z-f_1)}z}J_0\left(\frac{h_{Iq}f_1\rho}{(z-f_1)}\right)
    \label{eq:psil1}
\end{equation}


When $\Psi_{l_1}$ reaches the second lens, at $z=f_1+f_2$, the field becomes:


\begin{equation}
\Psi(\rho,z=f_1+f_2)=-e^{ik(f_1+f_2)}\frac{f_1}{f_2}\sum_{q=-N}^{N}A_{q}e^{i\frac{k\rho^2}{2f_2}}e^{i\frac{h_{Iq}^2f_1(f_1+f_2)}{2kf_2}}J_0\left(\frac{h_{Iq}f_1\rho}{f_2} \right)
\label{eq:exicitacao}
\end{equation}

At this point,  we have that (\ref{eq:exicitacao}) is the wave incident on the second lens, and that the field following it, $\Psi_{T}$, can also be calculated by using the Fresnel Diffraction Integral $\eqref{eq:fresnel}$, taking into account the transfer function of such lens:


\begin{equation}
  \Psi_T(\rho,z)=e^{ikz}\frac{-ik}{z}e^{\frac{ik\rho^2}{2z}}\sum_{q=-N}^{N}\int^\infty_0d\rho'\rho'(-1)
  e^{ik(f_1+f_2)}e^{i\frac{h_I^2f_1(f_1+f_2)}{2kf_2}}\frac{f_1}{f_2} A_{q} J_0\left(\frac{h_If_1\rho'}{f_2}\right)
  e^{i\frac{k\rho'^2}{2z}}J_0\left(\frac{k\rho\rho'}{z}\right)
  \label{eq:psit2}
\end{equation}
where $z=0$ is now the position of the second lens. By using eq.(\ref{eq:sol_int}), we get


\begin{equation}
\Psi_T(\rho,z)=(-1)e^{ik(f_1+f_2)}\frac{f_1}{f_2}e^{ikz}\sum_{q=-N}^{N}A_q e^{\frac{ih_{Iq}^2f_1(f_1+f_2)}{2kf_2}}e^{-i\frac{\left(f_1/f_2h_{Iq}\right)^2}{2k}z}J_0\left(\frac{h_{Iq} f_1\rho}{f_2}\right)
\label{psiT}
\end{equation}


Now, since
\begin{equation}
\frac{h_{Iq}^2}{2k}=k-\beta_{Iq}
\label{eq:betai_aux}
\end{equation}
and
\begin{equation}
\beta_{Iq}=Q+\frac{2\pi q}{L} \,\, ,
\label{eq:betai}
\end{equation}
the transmitted field becomes


\begin{equation}
    \Psi_T(\rho,z)=(-1)\frac{f_1}{f_2}e^{ik(f_1+f_2)}
    e^{iz\left[k-\left[\left(\frac{f_1}{f_2}\right)^2(k-Q)\right]\right]}
    \sum_{q=-N}^{N} A_q e^{i\frac{h_{Iq}^2f_1(f_1+f_2)}{2kf_2}} J_0\left(\frac{h_{Iq}f_1\rho}{f_2}\right)e^{iz\frac{2\pi q}{L\left(f_2/f_1\right)^2}}
    \label{FW2ln}
\end{equation}

In Equation \eqref{FW2ln}, it can be seen that appearance of the term $\exp{\left[\frac{ih_i^2f_1(f_1+f_2)}{(2kf_2)}\right]}$ into the summation. This exponential term affects the longitudinal pattern by displacing it. By making $z\rightarrow z+\frac{f_2}{f_1}(f_1+f_2)$, the field becomes



\begin{equation}
    \Psi_T(\rho,z)=(-1)\frac{f_1}{f_2}e^{ik(f_1+f_2)}e^{ik\left[\left(\frac{f_2}{f_1}\right)
    (f_1+f_2)\right]}e^{iz\left[k-\left[\left(\frac{f_1}{f_2}\right)^2(k-Q)\right]\right]}
    \sum_{q=-N}^{N} A_q J_0\left(\frac{h_{Iq}f_1}{f_2}\rho\right)e^{iz\frac{2\pi q}{L\left(f_2/f_1\right)^2}}
    \label{FW2ln2}
\end{equation}

 \noindent with

 \begin{equation}
    A_q=\frac{1}{L(f_2/f_1)^2}\int_{0}^{L(f_2/f_1)^2}F\left[\left(\frac{f_2}{f_1}\right)^2Z\right]e^{-i\frac{2\pi q}{L(f_2/f_1)^2}Z}dZ
    \label{eq:b}
 \end{equation}

  \noindent where $z=Z(f_2/f_1)^2$.
  
Equations (\ref{FW2ln2},\ref{eq:b}) show us that transverse spot size of such the FW scales with the lens focal ratio, $f_2 / f_1$, and the longitudinal intensity pattern scales with the quadratic factor $ (f_2 /f_1)^2$. 

\section{Conclusions}
In this present work, we analytically showed that, when a {\em Frozen Wave} propagates through a two convex lens system, the beam that emanates from the second lens is again a {\em Frozen Wave}, which however suffered important alterations: While its transverse spot size is scaled by the lens focal ratio, $f_2 / f_1$, its longitudinal pattern, which can be chosen on demand, is scaled by the quadratic factor $ (f_2 /f_1)^2$. The results here presented can be useful in the generation of structured light beams within very small spatial regions. [P.S.: the co-author E.R. thanks the Brazilian Fapesp for the PVE fellowship no.2019/12329-3.]

\end{document}